\documentclass[aps,twocolumn,groupedaddress,showpacs]{revtex4}
\usepackage{graphicx}

\begin{document}

\title[]
{Phase transition of clock models on hyperbolic lattice studied by corner
transfer matrix renormalization group method}

\author{A.~Gendiar$^1$, R.~Krcmar$^1$, K.~Ueda$^2$ and T.~Nishino$^2$}
\affiliation{$^1$Institute of Electrical Engineering, Centre of Excellence CENG,
Slovak Academy of Sciences, D\'{u}bravsk\'{a} cesta 9, SK-841~04, Bratislava, Slovakia\\
$^2$Department of Physics, Graduate School of Science, Kobe University,
Kobe 657-8501, Japan}

\date{\today}

\begin{abstract}
Two-dimensional ferromagnetic $N$-state clock models are studied on a 
hyperbolic lattice represented by tessellation of pentagons. 
The lattice lies on the hyperbolic plane with a constant negative scalar 
curvature. We observe the spontaneous magnetization, the internal energy, 
and the specific heat at the center of sufficiently large 
systems, where the fixed boundary conditions are imposed, for the cases $N\ge3$ 
up to $N = 30$. The model with $N = 3$, which is equivalent to the 3-state 
Potts model on the hyperbolic lattice, exhibits the first order phase
transition. A mean-field like phase transition of the second order is
observed for the cases $N\ge4$. When $N\ge5$ we observe the Schottky
type specific heat below the transition temperature, where its peak hight at
low temperatures scales as $N^{-2}_{~}$. From these facts we conclude
that the phase transition of classical XY-model deep inside the hyperbolic
lattices is not of the Berezinskii-Kosterlitz-Thouless type.
\end{abstract}


\maketitle

\section{Introduction}

Two-dimensional (2D) lattice models with continuous local spin symmetry,
such as the classical XY-model and the classical Heisenberg model on the
square lattice, do not have finite magnetization when temperature is finite.
This fact proved by Mermin and Wagner~\cite{Mermin} does not exclude
the presence of phase transition of the Berezinskii-Kosterlitz-Thouless (BKT)
type~\cite{Ber,KT}. These well-known facts are based on analysis in the
flat 2D plane.

Quite recently, Baek {\it et al.} studied the XY model on the heptagonal
lattice~\cite{Baek}, which is one of the hyperbolic lattices constructed
as a tessellation of heptagons on the hyperbolic plane, i.e., the 2D space 
with a constant negative curvature~\cite{Sausset}. 
By way of the Monte Carlo (MC) simulations for
open boundary systems, they concluded the absence of phase transition,
including that of the BKT type. Their result is in accordance with the 
thermodynamic property of the Ising model on the hyperbolic lattice, where
there is no singularity in the specific heat as shown by d'Auriac 
{\it et al.}~\cite{dAuriac}. These observations on the hyperbolic lattice
can be explained by the non-negligible effect of the 
system boundary~\cite{Chris1,Chris2}, 
which always has a finite portion of the system regardless of the
system size. 

It should be noted, as pointed by d'Auriac {\it et al.}, that
the presence of the ordered phase is not excluded in the region far from
the boundary~\cite{dAuriac}, although the area of such an ordered region
is negligibly small compared with the whole system on the hyperbolic lattice. 
The situation is similar to that of the statistical models on the Cayley 
tree, where its deep inside can be regarded as the Bethe lattice~\cite{Baxter}. 
Shima {\it et al.} studied
the Ising model on the hyperbolic lattice by the MC simulations, and
observed the mean-field like phase transition deep inside the
system~\cite{Shima,Hasegawa}. The mean field behavior is in accordance
with theoretical studies of phase transition in the infinitely large
hyperbolic lattices~\cite{Rietman,Doyon}. It can be expected that such
an order also appears in the case of the XY-model and the clock models.

In this paper we study $N(\ge3)$-state clock models  on the pentagonal
lattice~\cite{tech} up to $N = 30$ by use of the CTMRG
method~\cite{Nishino1,Nishino2,Nishino3}
modified for systems on the hyperbolic lattices~\cite{Ueda,Roman}.
The internal energy and the 
spontaneous magnetization {\it at the center of sufficiently large 
systems} are calculated numerically. 
In order to judge the presence of an ordered state deep inside the system,
we impose the ferromagnetic boundary conditions at the beginning of the 
iterative calculation of the CTMRG method. As we show in the following, the 
obtained results support the existence of the mean-field like phase 
transition for all the $N$ even in the limit $N \rightarrow \infty$, where 
the system coincides with the classical XY model. 

In the next section we introduce geometry of the pentagonal lattice and
consider the $N$-state clock model on it. A brief explanation of the CTMRG
method is presented. In Sec.~III we show numerical results on the spontaneous
magnetization, the internal energy, and the specific heat. We summarize
the observed phase transition.

\section{Clock models on pentagonal lattice}

\begin{figure}[tb]
\includegraphics[width=50mm,clip]{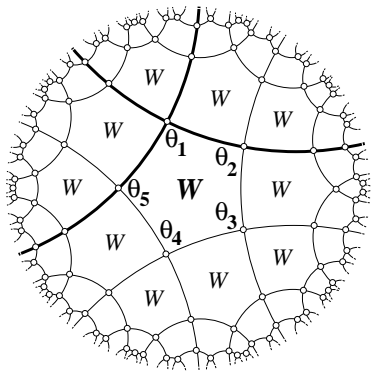}
\caption{The pentagonal lattice drawn in the Poincar\'{e} disc. The open
circles represent the $N$-state spin variables $\theta_i^{~}$. Two
geodesics drawn by thick arcs divide the system into four equivalent
quadrants.}
\label{f1}
\end{figure}

We consider the 2D lattice shown in Fig.~\ref{f1}, which is a tessellation of
regular pentagons. The lattice is in a curved plane with a constant negative
scalar curvature. Therefore, the Hausdorff dimension of the lattice is infinite.
For a technical reason in the CTMRG method, we have chosen the lattice with 
the coordination number four~\cite{tech}. Two geodesics drawn by the
thick arcs cross one another at a site labeled by $\theta_1^{~}$. By these
two arcs the whole lattice is divided into four equivalent parts called the
quadrants or the corners.

Let us introduce the $N$-state clock model on the pentagonal lattice.
On each lattice site there is an $N$-state spin variable $\theta_i^{~}$
where $i$ is the site index. The possible values of $\theta_i^{~}$ are
$2 \pi \xi / N$ with $\xi=0,1,2,\dots,N-1$. We consider the {\it angle}
$\theta_i^{~}$ as the internal degree of freedom. Therefore,
$\theta_i^{~}$ has nothing to do with the lattice geometry.
If there are only ferromagnetic interactions between neighboring
spin pairs, the Hamiltonian of the $N$-state clock model is written as
\begin{equation}
{\cal H}
= - J \sum_{\langle i j \rangle}^{~} \cos\left( \theta_i^{~}
- \theta_{j}^{~} \right) \, ,
\label{eq1}
\end{equation}
where $J > 0$ is the coupling constant. The summation
runs over all the nearest-neighbor pairs $\langle i j \rangle$.
The case $N = 2$ is nothing but the Ising model with coupling
interaction $J$ and this case has been studied~\cite{Ueda,Roman}.
The case $N = 4$ can be reduced to the Ising model with the coupling
$J / 2$. We thus chiefly discuss the case $N = 3$, which
is equivalent to the 3-state Potts model, and the cases 
$N \ge 5$ in the following.
In order to observe the phase transition deep inside
the system, we impose the ferromagnetic boundary conditions so that
all the spin variables at the system boundary are aligned in the direction
$\theta = 0$.

For convenience we represent this clock model as a special case of the
interaction-round-a-face (IRF) model on the hyperbolic lattice. For instance,
let us label the spins around a pentagon as shown in Fig.~\ref{f1}. The IRF
weight $W$, which is the local Boltzmann weight corresponding to this
pentagon, is obtained as
\begin{equation}
W(\theta_1^{~} \, \theta_2^{~} \, \theta_3^{~} \, \theta_4^{~}\, \theta_5^{~}) =
\prod\limits_{i=1}^5 \exp\left\{ \frac{J\cos\left( \theta_i^{~} 
- \theta_{i+1}^{~} \right)} {2 \, k_{\rm B}T} \right\} \, ,
\label{eq2}
\end{equation}
where $\theta_6\equiv\theta_1$. Having the IRF weight $W$ thus defined,
we can express the partition function of the whole system
\begin{equation}
{\cal Z} = \sum_{\{\theta\}}\prod \, W \, ,
\label{epf}
\end{equation}
where the product is taken for all the IRF weights in the pentagonal
lattice. The sum $\sum_{\{ \theta \}}^{~}$ is taken over all spin
configurations. 

In order to discuss the phase transition on the hyperbolic lattice,
let us consider a system whose size (or diameter) $L$ is 
far larger than the correlation 
length $\xi$. We divide the system into two parts, the boundary
area (BA) and the deep inside area (DIA). The former, BA,  is a ring-shaped area,
where all the sites in the area are within the distance of the order 
of $\xi$ from the system boundary. The latter, DIA, is the rest of the system, 
which we analyze in the following. Because of the hyperbolic geometry,
the portion of the BA with respect to the whole system is always finite even 
in the limit $L \rightarrow \infty$. The situation is similar to that of the Cayley 
tree~\cite{Baxter}. Thus the thermodynamic property of the whole system is 
always affected by the boundary condition, especially in low 
temperature~\cite{Chris1,Chris2}.
When $\xi$ is finite, it is possible to consider the thermodynamics of the
DIA, discarding the thermodynamic contribution from the BA, since we have 
assumed $L \gg \xi$ and therefore the size of the DIA is sufficiently large. 
When we collect numerical data of the DIA, we always treat sufficiently large 
systems that satisfy $L \gg \xi$, choosing such temperatures for which $\xi$
is at most of the order of 1000. We then detect the phase transition in the
DIA by extrapolation from both low- and high-temperature sides. 

We introduce Baxter's corner transfer matrix (CTM) $C$,
which represents the Boltzmann weight of a quadrant of the system~\cite{Baxter}.
The partition function ${\cal Z}$ is then expressed as ${\rm Tr}\  C^4$,
i.e., as the trace of the {\it density matrix} $\rho = C^4_{~}$.
Applying the concept of the density matrix renormalization~\cite{White1,White2,Sch},
a precise approximation of ${\cal Z}$ can be obtained for large scale systems 
by way of iterative numerical calculations~\cite{Nishino1,Nishino2,Nishino3}. 
These are the outline
of the CTMRG method, which can be applied to statistical models on
hyperbolic lattices~\cite{Ueda,Roman}. 

After we obtain the density matrix
$\rho$ for a sufficiently large system, we can calculate the expectation
values at the center of the system, which represent the thermodynamics
deep inside the system. For example, we can obtain the 
spontaneous magnetization
\begin{equation}
{\cal M}^{(N)}_{~} = {\rm Tr}\, \left[ \, \cos(\theta_c^{~}) \, 
\rho \, \right] \, / \, {\rm Tr}\, \rho \, ,
\end{equation}
where $\theta_{\rm c}^{~}$ represents the spin at the center of the system, and
the internal energy per bond
\begin{equation}
{\cal E}^{(N)}_{~} = -J \, {\rm Tr}\, \left [\, \cos(\theta_{\rm c}^{~}-\theta'_{\rm c}) 
\, \rho \, \right] \, / \, {\rm Tr}\,\rho \, ,
\end{equation}
where $\theta'_{\rm c}$ is the neighboring spin next to $\theta_{\rm c}^{~}$.
The specific heat ${\cal C}^{(N)}_{~}$ can be obtained by taking the numerical
derivative of ${\cal E}^{(N)}_{~}$ with respect to temperature $T$. It should be
noted that ${\cal M}^{(N)}_{~}$, ${\cal E}^{(N)}_{~}$, and ${\cal E}^{(N)}_{~}$
are not thermodynamic functions of the whole system but are those of the
area deep inside the system.

It has been known that the decay of the density matrix eigenvalues is very
fast for models on the hyperbolic lattices~\cite{Ueda,Roman}. The clock model
under study has the same feature in common. Therefore,
it is sufficient to keep a very small number of the degree of freedom
for the block spin variable $m$ in the formalism of CTMRG. Typically,
we keep $m\approx 2N$ states. We checked that further
increase of $m$ does not improve numerical precision in ${\cal M}^{(N)}_{~}$
and ${\cal E}^{(N)}_{~}$ any more, even at the vicinity of the phase transition.

\section{Numerical results}

\begin{figure}[tb]
\includegraphics[width=0.95\columnwidth,clip]{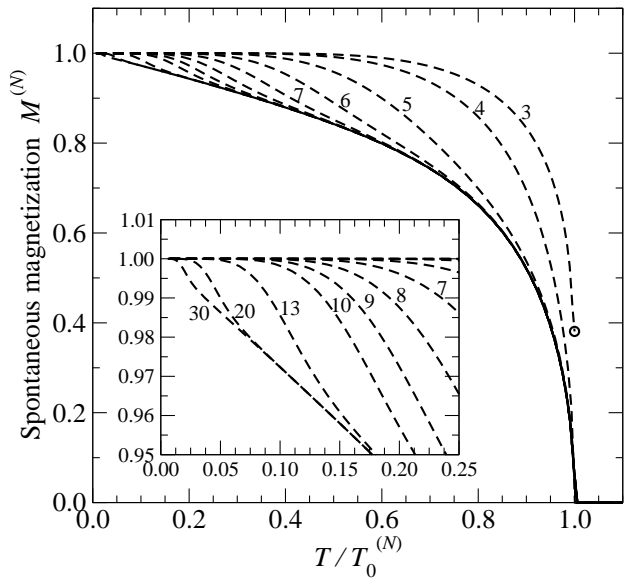}
\caption{Temperature dependence of the spontaneous
magnetization ${\cal M}^{(N)}_{~}$ for $3 \le N \le 30$. The
open circle denotes the discontinuity in ${\cal M}^{(3)}_{~}$.}
\label{f2}
\end{figure}

Throughout this section, we take the coupling constant $J$ in Eq.~(\ref{eq1})
to the unit of energy. For all the cases $N \ge 2$, we observe phase transition,
where the transition temperatures $T_{\rm 0}^{(N)}$ are listed in Table~\ref{t1}.
Note that $T_{\rm 0}^{(N)}$ converges to $T_{\rm 0}^{(\infty)}$ very fast
with respect to $N$.

\begin{figure}[tb]
\includegraphics[width=0.95\columnwidth,clip]{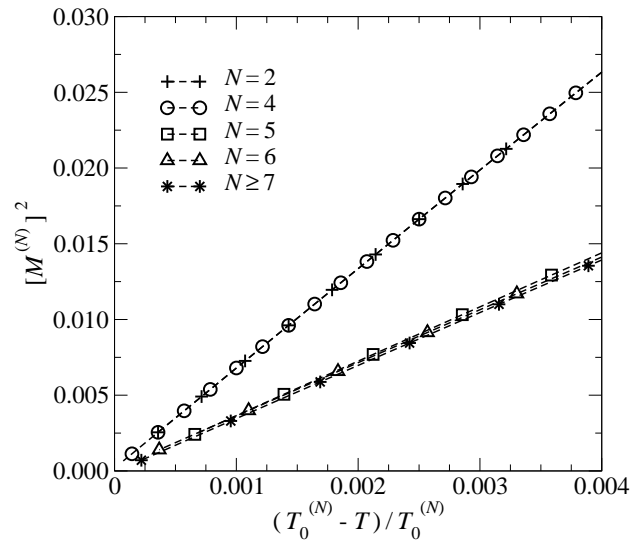}
\caption{Square of ${\cal M}^{(N)}_{~}$ with respect to $( T_0^{(N)} - T ) / T_0^{(N)}$.}
\label{f3}
\end{figure}

Figure~\ref{f2} shows the spontaneous magnetization ${\cal M}^{(N)}_{~}$
with respect to the rescaled temperature $T / T_{\rm 0}^{(N)}$.
(Under this rescaling, ${\cal M}^{(2)}_{~}$ and  ${\cal M}^{(4)}_{~}$
are identical.) If $N = 3$, the magnetization is discontinuous at
$T_{\rm 0}^{(3)}$. The 3-state clock model, which is equivalent to the
3-state Potts model, exhibits the first order phase transition if the
system is on the pentagonal lattice. 
This is a kind of mean-field behavior, since it is well
known that the mean-field approximation applied to the 3-state Potts
model on 2D lattices show the first order phase transition~\cite{Wu}.
In the vicinity of $T_0^{(N)}$  the magnetization ${\cal M}^{(N)}_{~}$
rapidly converges to the large $N$ limit ${\cal M}^{(\infty)}_{~}$.
The inset of Fig.~\ref{f2} displays the low-temperature behavior of
${\cal M}^{(N)}_{~}$ in details. Note that in the limit $N\to\infty$ the
magnetization ${\cal M}^{(N)}_{~}$ decreases linearly with $T$ at very
low temperatures.
Figure~\ref{f3} shows the square of ${\cal M}^{(N)}_{~}$ with respect to
$t = ( T_0^{(N)} - T ) / T_0^{(N)}$ for the cases other than $N = 3$.
It is obvious that the scaling relation ${\cal M}^{(N)}_{~} \propto t^{\beta}$
is satisfied with the exponent $\beta = \frac{1}{2}$.

\begin{table}[tb]
\caption {The transition temperatures $T_{\rm 0}^{(N)}$, the
critical exponents $\beta$, and positions of the specific
heat maximum $T_{\rm Sch}^{(N)}$.}
\label{t1}
\begin{ruledtabular}
\begin{tabular*}{\hsize}{
c@{\extracolsep{0ptplus1fil}}c@{\extracolsep{0ptplus1fil}}
c@{\extracolsep{0ptplus1fil}}c@{\extracolsep{0ptplus1fil}}
c@{\extracolsep{0ptplus1fil}}}
$N$-clock & $ T_{\rm 0}^{(N)}$ & $\beta$ &
$T_{\rm Sch}^{(N)}$  \\
\colrule
  2 & 2.7991 &  0.5  & ---            \\
  3 & 1.6817 &  ---    & ---            \\
  4 & 1.3995 &  0.5  & ---            \\
  5 & 1.3659 &  0.5  & ---            \\
  6 & 1.3625 &  0.5  & 0.62948 \\
  7 & 1.3623 &  0.5  & 0.46295 \\
  8 & 1.3622 &  0.5  & 0.35676 \\
  9 & 1.3622 &  0.5  & 0.28357 \\
10 & 1.3622 &  0.5  & 0.22997 \\
13 & 1.3622 &  0.5  & 0.13761 \\
20 & 1.3622 &  0.5  & 0.05864 \\
30 & 1.3622 &  0.5  & 0.02600 \\
\end{tabular*}
\end{ruledtabular}
\end{table}

\begin{figure}[tb]
\includegraphics[width=0.95\columnwidth,clip]{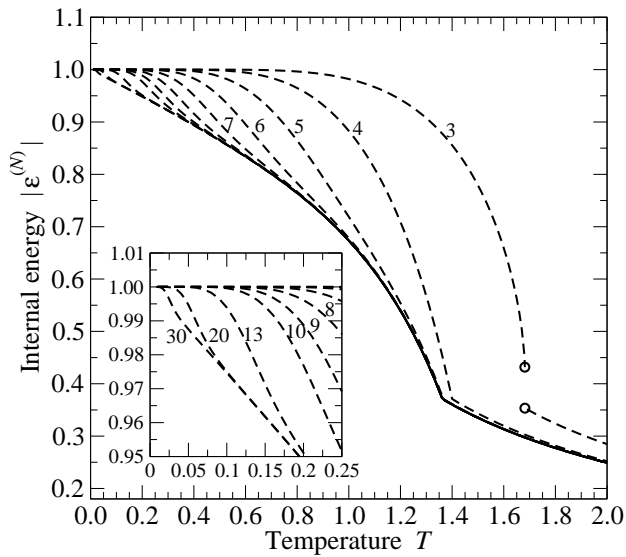}
\caption{The absolute value of the internal energy $| {\cal E}^{(N)}_{~} |$.
The open circles denote the jump in the case $N = 3$.}
\label{f4}
\end{figure}

Figure \ref{f4} shows the internal energy ${\cal E}^{(N)}_{~}$. There is a
finite jump in ${\cal E}^{(3)}_{~}$ at $T_0^{(3)}$, where the latent heat per
bond ${\cal L} = {\cal E}^{(3)}_{+} - {\cal E}^{(3)}_{-}$ is  $0.078$.
Analogously to the magnetization ${\cal M}^{(N)}_{~}$, the ${\cal E}^{(N)}_{~}$ is
linear in $T$ at low-temperature region in the limit $N\to\infty$.

\begin{figure}[tb]
\includegraphics[width=0.95\columnwidth,clip]{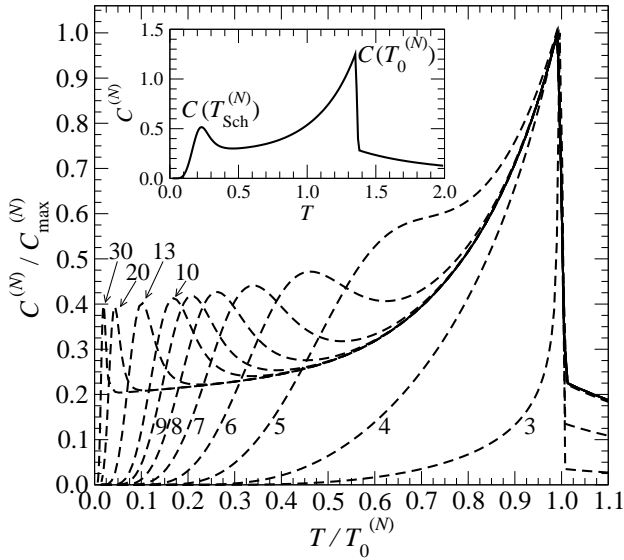}
\caption{The rescaled specific heat $C^{(N)}_{~} / C^{(N)}_{\rm ~max} $
versus the rescaled temperature $T / T_0^{(N)}$. The inset shows
a typical example for the case $N=10$ without rescaling.}
\label{f5}
\end{figure}

Figure~\ref{f5} shows the rescaled specific heat $C^{(N)}_{~} / C^{(N)}_{\rm ~max}$,
where $C^{(N)}_{\rm ~max}$ is the specific heat at $T_0^{(N)}$,
with respect to the rescaled temperature $T / T_0^{(N)}$. Evidently,
a discontinuity in the specific heat  is observed for the cases $N = 2$ and 
$N \ge 4$. Thus, the second order phase transition has the mean-field nature.
There is no indication of the BKT transition that is observed for
clock models on flat 2D lattices~\cite{Landau}.

\begin{figure}[t]
\includegraphics[width=0.95\columnwidth,clip]{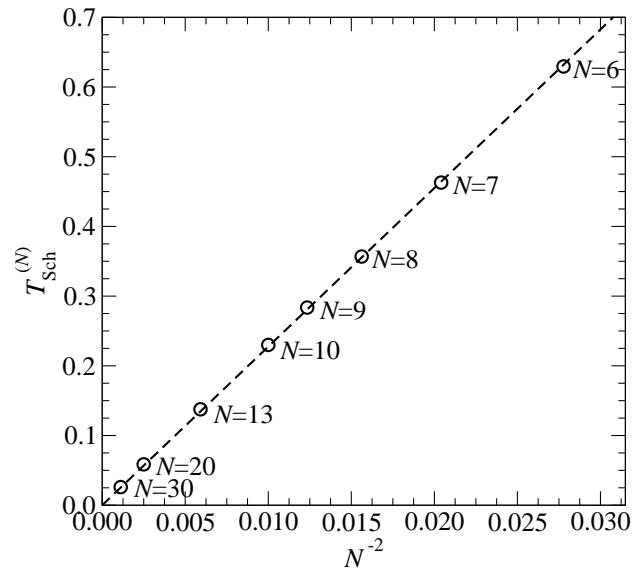}
\caption{The Schottky peak position $T_{\rm Sch}^{(N)}$
versus $1/{N^2_{~}}$.}
\label{f6}
\end{figure}

When $N$ is larger than 5, we observe the Schottky type peak
in the specific heat. Figure~\ref{f6} shows the $N$ dependence of
the Schottky peak position $T_{\rm Sch}^{(N)}$. As it is shown,
$T_{\rm Sch}^{(N)}$ is proportional to $1 / N^2_{~}$. This is
qualitatively in accordance with the energy scale of local excitation
$2 ( 2\pi / N )^2_{~} J$ from the completely ordered state.
It is thus concluded that the Schottky peak disappears in the limit
$N \rightarrow \infty$ and that the specific heat of the classical XY
model on the pentagonal lattice remains finite even at $T = 0$.

\section{Conclusions}

We have studied the $N$-state clock models on the pentagonal lattice,
which is a typical example of the hyperbolic lattices. The phase transition
deep inside the system is observed by use of the CTMRG method.
From the critical exponent $\beta = \frac{1}{2}$ for the spontaneous 
magnetization and the jump in the specific heat,
we conclude that the phase transition for $N = 2$ and $N \ge 4$
is mean-field like, provided that the ferromagnetic boundary conditions
are imposed. The Hausdorff dimension, which is infinite for the
hyperbolic lattices, is essential in the observed critical behavior. We
conjecture that the phase transition deep inside the system is also present 
for systems with free boundary conditions. 

In the case when $N = 3$, where
the system is equivalent to the 3-state Potts model,
we observed the first-order phase transition. 
Since the $q$-state Potts model tends to exhibit the first-order transition for
larger $q$~\cite{Wu}, it is expected that the transition of $q \ge 3$ Potts
models on the pentagonal lattice is of the first order. We have partially 
confirmed the behavior for several values of $q$ and we
conjecture that the transition is of the first order on any kind of
hyperbolic lattices when $q \ge 3$.

We observed stable ferromagnetic states below $T_0^{(N)}$ even in the
continuous limit $N \rightarrow \infty$. This fact does not contradict to the
Mermin-Wagner theorem~\cite{Mermin} since the pentagonal lattice is
not on the flat 2D plane. The vortex energy on hyperbolic lattices might
be larger than that on the flat lattice. The difference may elucidate the
absence of the BKT phase transition on the pentagonal lattice.

\section*{Acknowledgments}

The Slovak Agency for Science and Research grant APVV-51-003505 and
Slovak VEGA grant No. 2/6101/27 are acknowledged (A.G. and R.K.). This
work is also partially supported by Grant-in-Aid for Scientific Research from
Japanese Ministry of Education, Culture, Sports, Science and Technology
(T.N. and A.G.).

\end{document}